\documentclass[12pt]{article}
\usepackage[pdftex,pagebackref,letterpaper=true,colorlinks=true,
            pdfpagemode=none,urlcolor=blue,linkcolor=blue,
            citecolor=blue,pdfstartview=FitH]  {hyperref}

\usepackage{amsmath,amsfonts}
\usepackage{graphicx}
\usepackage{wrapfig}
\usepackage{color}

\begin{document}

\begin{center}
\begin{LARGE}
{\bf Richardson's Forecast:}

\smallskip
{\bf the Dream and the Fantasy}
\end{LARGE}

\begin{large}
Peter Lynch%
\footnote{Peter.Lynch@ucd.ie} \\
School of Mathematics and Statistics, University College Dublin.
\end{large}

\end{center}


\bigskip
\begin{small}
\begin{quote}
{Perhaps, some day in the dim future it will be possible to advance
the computations faster than the weather advances \dots\ . But that is a dream [WPNP, p. vii].
}
\end{quote}
\end{small}

\subsubsection*{Introduction}

A remarkable book on weather forecasting was published just one
hundred years ago.  Written by the brilliant and prescient applied
mathematician, Lewis Fry Richardson, \emph{Weather Prediction by
Numerical Process} was published by Cambridge University Press and
went on sale in 1922 at a cost of 30 shillings (\pounds 1.50).  With
a print run of just 750 copies, it was not a commercial success and
was still in print thirty years after publication.  It was re-issued
in 1965 as a Dover paperback. Cambridge University Press reprinted
the book in 2007, with a foreword by Peter Lynch. Described as a
second edition, it differs in no essential way from the 1922 edition.

\emph{Weather Prediction by Numerical Process} (denoted WPNP) is a
strikingly original scientific work, one of the most remarkable books 
on meteorology ever written.  In it, Richardson described a systematic mathematical
method for predicting the weather and demonstrated its application by carrying out 
a trial forecast. Richardson's innovative approach was fundamentally sound,
but the method devised by him was utterly impractical at the time of its 
publication and the results of his trial forecast appeared to be
little short of outlandish. As a result, his ideas were eclipsed for decades.
For a brief biographical sketch of Richardson, see the \emph{Benchmark} article
by Lea (2012) and for a full biography, see Ashford (1985).

\subsubsection*{Background to Scientific Forecasting}

The basic ideas of numerical forecasting and climate modelling were
developed more than a century ago, long before the first electronic
computer was constructed. American meteorologist Cleveland Abbe and
Norwegian physicist Vilhelm Bjerknes recognized that predictions
of meteorological phenomena could be based on the application of
hydrodynamics and thermodynamics to the atmosphere (Abbe 1901;
Bjerknes 1904). They each specified the required process: analysis
of the initial state and time integration of the governing equations
from that state.  They listed the system of mathematical equations
needed and outlined the steps required to produce a forecast.
Realizing the numerous practical difficulties, neither Abbe nor
Bjerknes attempted to implement their techniques.

A first tentative trial to produce a forecast using the laws of
physics was made by Felix Exner (1908), working in Vienna. His
efforts yielded a realistic forecast for the particular case that
he chose. Despite the restricted applicability of his technique,
his work was a first attempt at systematic, scientific weather
forecasting. However, it involved some drastic simplifications and
was far from providing anything of use for practical forecasting.

\subsubsection*{\emph{Weather Prediction by Numerical Process}}

Lewis Fry Richardson first learned of Bjerknes' plan for rational
forecasting in 1913, when he took up employment with the Meteorological
Office. Richardson's forecasting scheme amounts to a precise and
detailed implementation of the prognostic component of Bjerknes'
programme. It is a highly intricate procedure: as Richardson observed,
``the scheme is complicated because the atmosphere is complicated.''
It also involved an enormous volume of numerical computation and
was quite impractical in the pre-computer era. But Richardson was
undaunted, expressing his dream that ``some day in the dim future
it will be possible to advance the computations faster than the
weather advances.''

Earlier, Richardson had applied an approximate numerical method to
the solution of partial differential equations to investigate the
stresses in masonry dams. He realized that this method had potential
for use in a wide range of problems. The idea of numerical weather
prediction appears to have germinated in his mind for several years.
Around 1911, Richardson had begun to think about the application
of his approach to the problem of forecasting the weather. He stated
in the Preface of WPNP that the idea ``grew out of a study of finite
differences and first took shape in 1911 as a fantasy.''
The fantasy was that of a forecast factory, which we will discuss below.

Upon joining the Met Office, Richardson was appointed Superintendent
of Eskdalemuir Observatory in the Southern Uplands of Scotland and
began serious work on numerical forecasting. In May 1916 he resigned
from the Met Office in order to work with the Friends Ambulance
Unit in France. By this time, he had completed the formulation of
his scheme and had set down the details in the first draft of his
book. But he was not concerned merely with theoretical rigour and
wished to include a fully worked example to demonstrate how the
method could be put to use.

Richardson assumed that the state of the atmosphere at any point
could be specified by seven numbers: pressure, temperature, density,
water content and velocity components eastward, northward and upward.
He formulated a description of atmospheric phenomena in terms of
seven partial differential equations. To solve them, he divided the
atmosphere into discrete columns of extent 3 degrees east-west and
200 km north-south, giving 120$\times$100 = 12,000 columns to cover
the globe. Each of these columns was divided vertically into five
cells. The values of the variables were given at the centre of each
cell, and the differential equations were approximated by expressing
them in finite difference form. The rates of change of the variables
could then be calculated by arithmetical means.  These rates enabled
Richardson to calculate the variables at a later time.

Richardson calculated the initial changes over a six hour period
in two columns over central Europe, one for mass variables and one
for winds.  This was the extent of his  forecast. In this trial
forecast, he calculated a change of atmospheric pressure, at a point
near Munich, of 145 hPa in 6 hours. This was a totally unrealistic
value, two orders of magnitude too large. The failure may be explained
in terms of atmospheric dynamics. We return to the cause of this
after first considering the reaction of other researchers to
Richardson's work.

\subsubsection*{Initial Reactions to WPNP}

The initial response to Richardson's book was unremarkable and must
have been disappointing to him.  WPNP was widely reviewed with
generally favourable comments, but the impracticality of the method
and the abysmal failure of the solitary sample forecast inevitably
attracted adverse criticism. Napier Shaw, reviewing the book for
Nature, wrote that Richardson ``presents to us a \emph{magnum opus}
on weather prediction.'' However, in reference to the trial forecast,
he observed that ``the wildest guess [at the pressure change] would
not have been wider of the mark.'' He also questioned Richardson's
conclusion that wind observations were the real cause of the error,
and his dismissal of the geostrophic wind.

Edgar W.~Woolard, later an editor of \emph{Monthly Weather Review},
wrote a positive review of the book for that journal, expressing
the hope that other investigators would be inspired by Richardson's
work to continue its development.  However, nobody else continued
working along his lines, perhaps because the forecast failure acted
as a  deterrent, perhaps because the book was so difficult to read,
with its encyclopaedic but distracting range of topics. Alexander
McAdie, Professor of Meteorology at Harvard, wrote ``It can have
but a limited number of readers and will probably be quickly placed
on a library shelf and allowed to rest undisturbed by most of those
who purchase a copy.'' Indeed, this is essentially what happened
to the book.

A most perceptive review by F.~J.~W.~Whipple of the Met Office came
closest to understanding Richardson's unrealistic forecast, postulating
that rapidly-travelling waves contributed to its failure: ``The
trouble that he meets is that quite small discrepancies in the
estimate of the strengths of the winds may lead to comparatively
large errors in the computed changes of pressure.''  Whipple appears
to have had a far clearer understanding than did Richardson himself
of the causes of the forecast catastrophe

Richardson's work was not taken seriously by most meteorologists
of the day and his book failed to have any significant impact on
the practice of weather forecasting during the decades following
its publication. Nevertheless, his work is the foundation upon which
modern forecasting is built.  Several key developments in the ensuing
decades set the scene for later progress. Timely observations of
the atmosphere in three dimensions were becoming available following
the invention of the radiosonde. Developments in the theory of
meteorology provided crucial understanding of atmospheric dynamics
and the filtered equations necessary to calculate the synoptic-scale
tendencies. Advances in numerical analysis led to the design of
stable algorithms. Finally, the development of digital computers
provided a way of attacking the enormous computational task involved
in weather forecasting, all leading to the first weather prediction
by computer (Charney \emph{et al.}, 1950).  The history leading to the
emergence of modern operational numerical weather prediction is
described in Lynch (2006).

\subsubsection*{Solution of the ``Noise Problem''}

In the atmosphere there are long-period variations dominated by the
effects of the Earth's rotation --- these are the meteorologically
significant rotational modes --- and short-period oscillations
called gravity waves, having speeds comparable to that of sound.
For many purposes, the gravity waves, which are normally of small
amplitude, may be disregarded as noise. However, they are solutions
of the governing equations and, if present with spuriously large
amplitudes in the initial data, can completely spoil the forecast.

The most obvious approach to circumventing the noise problem is to
construct a forecast by combining many time steps which are short
enough to enable accurate simulation of the detailed high frequency
variations.  If Richardson had extended his calculations, taking a
large number of very small time steps, his results would have been
noisy, but the mean values would have been meteorologically reasonable.
Of course, the attendant computational burden made this utterly
impossible for Richardson.

A more practical approach to solving the problem is to modify the
governing equations so that the gravity waves no longer occur as
solutions. This process is known as filtering the equations.  The
first successful computer forecasts (Charney, \emph{et al.}, 1950) were
made with the barotropic vorticity equation, which has low frequency
but no high frequency solutions. Another approach is to adjust the
initial data so as to reduce or eliminate the gravity wave components.
The adjustments can be small in amplitude but large in effect.  This
process is called initialization, and it may be regarded as a special
form of smoothing.

Lynch (2006) showed that the digital filtering initialization method
yields realistic tendencies when applied to Richardson's data. He
found that when appropriate smoothing was applied to the initial
data, using a simple digital filter, the initial tendency of surface
pressure was reduced from the unrealistic 145 hPa in 6 hours to a
reasonable value of less than 1 hPa in 6 hours. The forecast was
in good agreement with the observed pressure change.  The rates of
change of temperature and wind were also realistic. This confirmed
the root cause of the failure of Richardson's trial forecast:
unavoidable errors in the analysed winds resulted in spuriously
large values of divergence, which caused the anomalous pressure
tendencies. In effect, the analysis contained gravity wave components
of unrealistically large amplitudes.

The absence of gravity waves from the initial data results in
reasonable initial rates of change, but it does not automatically
allow the use of large time steps. The existence of high frequency
solutions of the governing equations imposes a severe restriction
on the size of the time step allowable if reasonable results are
to be obtained. The restriction, known as the CFL criterion, can
be circumvented by treating those terms of the equations that govern
gravity waves in a numerically implicit manner; this distorts the
structure of the gravity waves but not of the low frequency modes.
In effect, implicit schemes slow down the faster waves, thus removing
the cause of numerical instability. Most modern forecasting models
avoid the pitfall that trapped Richardson by means of initialization
followed by semi-implicit integration.

\subsubsection*{Richardson's Later Work}

After the First World War, Richardson's research focussed primarily
on atmospheric turbulence. He had encapsulated the essence of the
cascade of turbulent energy in a simple and oft-quoted rhyme embedded
in the text of WPNP: \emph{Big whirls have little whirls that feed
on their velocity, and little whirls have lesser whirls and so on
to viscosity}.  Richardson's dense writing style is occasionally
lightened in this way by a whimsical touch as, when discussing the
tendency of turbulence to increase diversity, he writes ``This one
can believe without the aid of mathematics, after watching the
process of stirring together water and lime-juice'' (WPNP, page
101).  Several of his publications during this period are still
cited by scientists. In one of the most important  --- \emph{The
supply of energy from and to atmospheric eddies} --- he derived a
criterion for the onset of turbulence, introducing what is now known
as the Richardson Number.

Around 1926, Richardson made a deliberate break with meteorological
research. He was distressed that his turbulence research was being
exploited for military purposes. Indeed, this knowledge impelled
him to destroy a large volume of his research papers. In a much
later study, Richardson investigated the separation of initially
proximate tracers in a turbulent flow and arrived empirically at
his ``four-thirds law'': the rate of diffusion is proportional to
the separation raised to the power 4/3. This was later established
more rigorously by Andrey  Kolmogorov using dimensional analysis.

\subsubsection*{Advances in Computing: From ENIAC to PHONIAC}

The first weather forecast (technically, a hindcast) made with a
digital computer was performed on the ENIAC (Electronic Numerical
Integrator and Computer) by a team of scientists at Princeton.  The
Princeton team were aware that Richardson's initial tendency field
was completely wrong because he was not able to evaluate the
divergence. They realised that a filtered system of equations would
have dramatic implications for numerical integration. It would
obviate the problem of gravity-wave noise and would permit a much
larger time step to be used. They integrated the barotropic vorticity
equation from real initial conditions and produced four realistic,
if far from perfect, forecasts. For a full account, see Chapter 10
of (Lynch, 2006).

It is gratifying that Richardson was made aware of the success in
Princeton; Jule Charney sent him a copy of the Tellus paper (Charney,
Fj\o rtoft and von Neumann, 1950). In his response, Richardson
congratulated Charney ``on the remarkable progress which has been
made in Princeton; and on the prospects for further improvement
which you indicate''.  He concluded by saying that the ENIAC results
were ``an enormous scientific advance'' on the single, and quite wrong,
forecast in which his own work had ended.

To illustrate the dramatic growth in computer power since the days
of the ENIAC, one of the forecasts was re-run on a small mobile
phone, a Nokia 6300, which had raw computational power comparable
to a CRAY-1, the first super-computer acquired by the European
Centre for Medium-Range Weather Forecasts (ECMWF). The computation
time for a 24-hour forecast on ENIAC was about 24 hours. The time
on the Nokia, christened Portable Hand Operated Numerical Integrator
and Computer (PHONIAC), was less than one second (Lynch \&\ Lynch,
2008).

\begin{figure}[t]
\begin{center}
\includegraphics[width=0.90\textwidth]{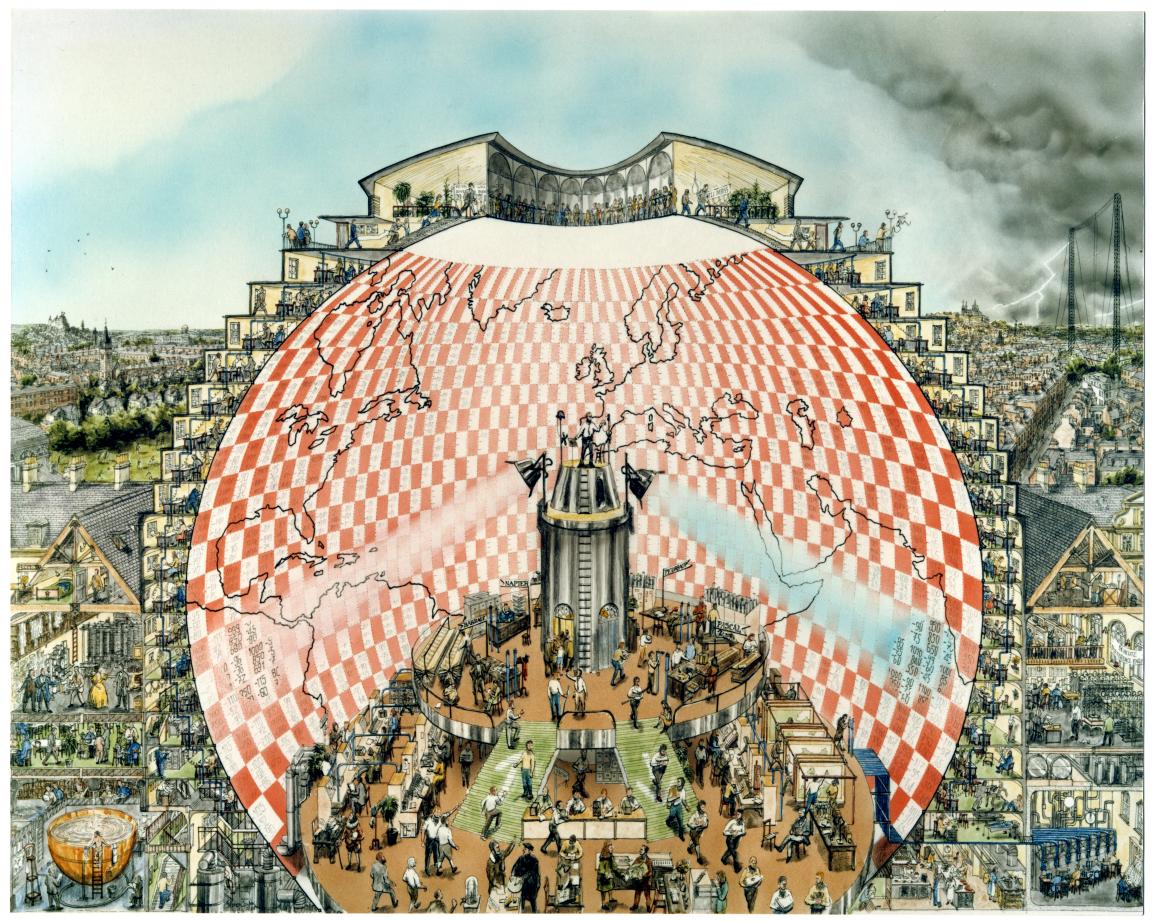}
\caption{Illustration of Richardson's forecast factory
\copyright Stephen Conlin (\url{www.pictu.co.uk}).}
\end{center}
\end{figure}

\subsubsection*{Richardson's Fantastic Forecast Factory}

The computation of his forecast was prodigious, taking Richardson
some two years to complete. How could the enormous number of
calculations necessary for a practical forecast ever be done?
Richardson estimated that it would require 64,000 people just to
keep up with the weather.  In WPNP, he described his fantasy: a
\emph{Forecast Factory} like a large theatre-in-the-round --- think
of the Royal Albert Hall --- a circular building with a great central
chamber, the walls painted to form a map of the globe. A large team
of (human) computers are busy within the building calculating the
future weather. The work is coordinated by a Director of Operations.
Standing on a central dais, he `conducts the computations' by
signalling with a spotlight to those who are racing ahead or
behindhand.

In 1986, an Irish artist, Stephen Conlin, created an illustration
of the forecast factory (Fig.~1). This painting, in ink and water
colours, is a remarkable work, replete with narrative details. The
painting depicts a large spherical building with a vast central
chamber. Four banners identify major pioneers of computing: John
Napier, Charles Babbage, George Boole and the first computer
programmer, Ada Lovelace. The painting is described in detail in
an article in Weather (Lynch, 2016).

There are surprising similarities between Richardson's forecast
factory and a modern massively parallel processor (MPP). Richardson
envisaged a large number of (human) computers working in synchrony
on different sub-tasks. In the fantasy, the forecasting job is
sub-divided using domain decomposition, a technique often used in
parallel computers today. Richardson's scheme involved nearest-neighbour
communication, analogous to message-passing techniques used in MPPs.
The man in the pulpit functioned like a synchronization and control
unit. Thus, the logical structures of the forecast factory and an
MPP have much in common.

\subsubsection*{Summary}

Richardson's dream was that scientific weather forecasting would
one day become a practical reality. Modern weather forecasts are
made by calculating solutions of the mathematical equations governing
the atmosphere. The solutions are generated by complex simulation
models implemented on powerful computer equipment. His dream has
indeed come true.

The development of comprehensive models of the atmosphere is
undoubtedly one of the finest achievements of meteorology in the
twentieth century. Numerical models continue to evolve, with
substantial  developments in data assimilation to produce improved
initial conditions, new numerical algorithms for more precise and
faster computations, and a probabilistic approach with ensemble
forecasts that quantify uncertainties in an operational environment.
These developments have made the dreams of Abbe, Bjerknes and
Richardson an everyday reality. Meteorology is now firmly established
as a quantitative science, and its value and validity are demonstrated
daily by the acid test of any science, its ability to predict the
future.


\subsubsection*{References}

[This small selection of reference may be supplemented by the extensive bibliography in Lynch (2006)]

\begin{itemize}

\item
Abbe, C., 1901: The physical basis of long-range weather forecasts. 
\emph{Mon.~Wea.~Rev.}, \textbf{29}, 551--561.

\item
Ashford, O.~M., 1985: \emph{Prophet --- or Professor?
The Life and Work of Lewis Fry Richardson}. Adam Hilger, xiv + 304 pp.

\item
Bjerknes, V., 1904: Das Problem der Wettervorhersage, betrachtet vom Standpunkte
der Mechanik und der Physik. [The problem of weather prediction, considered from 
the viewpoints of mechanics and physics.] \emph{Meteor.~Z.}, \textbf{21}, 1--7
[translated and edited by E.~Volken and S.~Br\"{o}nnimann, 2009: \emph{Meteor.~Z.},
\textbf{18}, 663--667].

\item
Charney, J.~G., R.~Fj\o rtoft, and J.~von Neumann, 1950:
Numerical integration of the barotropic vorticity equation. \emph{Tellus}, \textbf{2}, 237--54.

\item
Lea, Chris, 2012: Lewis Fry Richardson: the Father of Weather Forecasting. 
\emph{Benchmark}, \textbf{12}, 36--38.

\item
Lynch, Peter, 2006:
\emph{The Emergence of Numerical Weather Prediction: Richardson's Dream.}
Cambridge University Press, 279 pp.

\item
Lynch, Peter, 2016: An artist's impression of Richardson's fantastic forecast factory.
\emph{Weather}, \textbf{71}, 14--18.

\item
Lynch, Peter and Owen Lynch, 2008: Forecasts by PHONIAC.
\emph{Weather}, \textbf{63}, 324--326.

\item
Richardson, L. F., 1922: \emph{Weather Prediction by Numerical Process.}
Cambridge University Press, xii + 236 pp.  Reprinted by Dover Publications, 1965,
with a new Introduction by Sydney Chapman, xvi + 236 pp.
Second ed., Cambridge University Press, 2007, with Foreword by Peter Lynch.

\end{itemize}

\end{document}